\documentclass[apjl]{emulateapj}
\pdfoutput=1
\usepackage{natbib}
\usepackage{amsmath}
\usepackage{graphicx}
\usepackage{epsfig}

\def\Dwa{$\,$\uppercase\expandafter{\romannumeral5}$\,$}

\def\sless{\lower2pt\hbox{$\buildrel {\scriptstyle <}
   \over {\scriptstyle\sim}$}}

\def\sgreat{\lower2pt\hbox{$\buildrel {\scriptstyle >}
   \over {\scriptstyle\sim}$}}

\begin{document}

\slugcomment{Received June 18, 2010; Accepted July 7, 2010}

\title{Dimension as a Key to the Neutrino Mechanism of Core-Collapse Supernova Explosions}

\author{J. Nordhaus\altaffilmark{1}, A.~Burrows\altaffilmark{1}, A. Almgren\altaffilmark{2}, J. Bell\altaffilmark{2}}
\altaffiltext{1}{Department of Astrophysical Sciences, Princeton University, 
Princeton, NJ 08544 USA; nordhaus,burrows@astro.princeton.edu}
\altaffiltext{2}{Computational Research Division, Lawrence Berkeley National Lab, Berkeley, CA 94720, USA; ASAlmgren@lbl.gov,JBBell@lbl.gov}

\begin{abstract}
We explore the dependence on spatial dimension 
of the viability of the neutrino heating mechanism of core-collapse 
supernova explosions. We find that the tendency to explode is a 
monotonically increasing function of dimension, with 3D requiring 
$\sim$40$-$50\% lower driving neutrino luminosity than 1D and $\sim$15$-$25\%
lower driving neutrino luminosity than 2D.  Moreover, we find that 
the delay to explosion for a given neutrino luminosity is 
always shorter in 3D than 2D, sometimes by many hundreds of 
milliseconds. The magnitude of this dimensional 
effect is much larger than the purported magnitude of a variety 
of other effects, such as nuclear burning, inelastic scattering, or general 
relativity, which are sometimes invoked to bridge the gap between
the current ambiguous and uncertain theoretical situation and the fact of
robust supernova explosions. Since real supernovae occur in three 
dimensions, our finding may be an important step towards unraveling one of the most problematic  
puzzles in stellar astrophysics. In addition, even though in 3D we
do see pre-explosion instabilities and blast asymmetries, unlike the 
situation in 2D, we do not see an obvious axially-symmetric dipolar shock oscillation.
Rather, the free energy available to power instabilites seems to be 
shared by more and more degrees of freedom as the dimension increases.
Hence, the strong dipolar axisymmetry seen in 2D and previously identified as 
a fundamental characteristic of the shock hydrodynamics 
may not survive in 3D as a prominent feature.
\end{abstract}


\section{Introduction}
\label{intro}

It was shown some time ago that multi-dimensional instabilities obtain 
and are probably central to the core-collapse supernova mechanism
(Burrows \& Fryxell 1992; Herant, Benz, \& Colgate 1992; Herant et al. 1994;
Burrows, Hayes, \& Fryxell 1995; Janka \& M\"uller 1996). However, 
it has not been clear whether modest effects, particularly in the 
neutrino sector, that individually might have little consequence,
could accumulate to collectively push the core beyond the threshold of
instability into explosion (Mezzacappa et al. 1998,2001,2007; Bruenn et 
al. 2007,2010; Marek \& Janka 2009). Moreover, though the explosion 
energy is only 10$^{51}$ ergs, since more than 10$^{53}$ ergs
of neutrinos issue from collapse, the supernova phenomenon
might seem to be a ``one-percent" effect, necessitating attention to 
every detail to see the qualitative effect that is the supernova.
However, in the crucial early post-bounce phase of a few hundred milliseconds,
only a few times 10$^{52}$ ergs in electron-type neutrinos emerges, making the neutrino-driven
explosion more like a ``tens of percent" effect, challenging the notion that fine detail
in the neutrino sector is the key to the viability of the neutrino explosion 
mechanism. 

It may well prove to be the case that the fundamental impediment
to progress in supernova theory over the last few decades has not been lack 
of physical detail, but lack of access to codes and computers with which to properly simulate
the collapse phenomenon in 3D. This could explain the agonizingly
slow march since the 1960's towards demonstrating a robust mechanism of explosion.
State-of-the-art two-dimensional simulations are still ambiguous and problematic (\S\ref{status}), though
they manifest instability and turbulence and increase the dwell time of
matter in the gain region (Bethe \& Wilson 1985), and, hence, the efficiency of the neutrino-matter
coupling (Murphy \& Burrows 2008).  The gain region is the low-optical
(-neutrino) depth region behind the shock where neutrino heating
exceeds neutrino cooling. Moreover, relative to the spherical case, 2D effects 
have been shown to enlarge the gain region itself and place 
more matter in shallower climes of the gravitational potential 
well (from which it is easier to launch).  Two-dimensional models also 
enable accretion in one quadrant to power explosion in another (Burrows et al. 
2007b), something impossible in 1D and one reason neutrino-driven
explosions in 1D are generally thwarted.  In 1D, an outward explosion diminishes
the mass accretion rate onto the residual core that powers a good fraction of the driving
neutrino luminosity during the critical incipient explosive phase.

However, the difference in the character of 3D turbulence, with its extra
degree of freedom and inverse cascade to smaller turbule scales than are found in 2D,
may further increase the time matter spends in the gain region, so that 
the critical condition for explosion (Burrows \& Goshy 1993; Murphy \& Burrows 2008)
is more easily achieved. Our thesis in embarking upon this project was that if 
going from 2D to 3D improves prospects for the neutrino-driven
explosion by the same degree as already demonstrated by going from 1D to 2D (see \S\ref{status}), 
then the neutrino-driven mechanism would perhaps be shown to be not only viable, but robust.
In that case, the details of neutrino interactions, general relativity, and nuclear physics
would be of secondary importance for demonstrating the mechanism of explosion, though would still be
centrally important for determining the actual explosion energies, the magnitude and systematics of pulsar kicks, 
the residual neutron star masses, and the nucleosynthesis, to name just a few 
important aspects of the core-collapse supernova phenomenon.

Hence, in this paper we have conducted a parameter study, similar to that performed 
by Murphy \& Burrows (2008) (see \S\ref{castro}), but directly comparing 1D (spherical), 
2D (axisymmetric), and 3D (Cartesian) simulations, while varying the 
driving neutrino luminosity.  To isolate the effect of dimension, 
we ensured that all other aspects of the runs (equation of state, progenitor model, 
and neutrino heating algorithm, etc.) were the same. Note that for a given massive-star progenitor, the 
accretion rate in front of the stalled shock is independent of dimension, since the initial
model is spherically symmetric and the post-shock matter is out of sonic contact with the supersonic infalling 
matter in front of the shock.  Therefore, the character of infall, the evolution of the accretion rate 
and the mass interior to the shock, and by extension the evolution of the neutrino luminosities, 
are functions more of progenitor than of dimension (until explosion).

In what follows, we repeat the 1D and 2D study of Murphy \& Burrows (2008),
but, extending it to 3D, verify that the viability of explosion   
by delayed neutrino heating is indeed a monotonically increasing function of dimension.
In going from 2D to 3D, the enhancements in the efficiency of the neutrino-matter coupling in 
the gain region and the decreases in the critical luminosity are comparable to 
those seen in going from 1D to 2D.  In a very real sense, this implies that it is 
$\sim$50\% easier to explode in 3D than in 1D, a huge difference.  
Since real supernovae occur in three domensions, this conclusion may be a major step
in unraveling one of the most recalcitrant puzzles in astrophysics. 

In \S\ref{status}, we review the current status of numerical supernova theory.
Then, in \S\ref{castro} we summarize the CASTRO code and the simple algorithm by which we incorporate 
neutrino heating and cooling for the purposes of this parameter study.  We follow in \S\ref{results} with a discussion of our
general results, highlighting the role of dimension alone in facilitating explosion. We
conclude in \S\ref{conclusions} with a summary of what we have learned.

\section{Summary of Current Simulation Efforts and Results}
\label{status}

Colgate, Grasberger, \& White (1961) pioneered the direct (hydrodynamic) mechanism of core-collapse supernovae,
in which the bounce shock, launched when collapse is reversed, propagated outward without stalling.
Following this, Colgate \& White (1966) proposed the neutrino mechanism, wherein 
the agency of explosion was a burst of neutrino heating behind the shock after bounce.
Neutrinos must be included at the high temperatures and densities achieved in collapse.  
Arnett (1966) and Wilson (1971) refined this model, with the latter challenging 
its efficacy. These simulations were done in spherical symmetry and employed less 
sophisticated numerical techniques and cruder microphysics than can be marshalled today.
Importantly, current thinking is that the direct hydrodynamic mechanism never works. 
The neutrino burst that occurs at shock breakout of the neutrinosphere and, to a lesser degree, 
photodissociation of nuclei into nucleons at the nascent shock debilitate it into accretion.
Moreover, and disappointingly, modern simulations with the latest nuclear and neutrino physics, up-to-date progenitor models, 
and sophisticated numerics demonstrate that if stellar core collapse is constrained to be spherically-symmetric 
most cores can not supernova (Burrows, Hayes, and Fryxell 1995; Mezzacappa et al. 2001;
Rampp and Janka 2000,2002; Liebend\"orfer et al. 2001,2005; Buras
et al. 2003; Thompson, Burrows, and Pinto 2003). 
However, Kitaura et al. (2006) do obtain a spherical neutrino-wind-driven explosion for 
the 8.8-M$_{\odot}$ progenitor of Nomoto \& Hashimoto (1988), after a slight post-bounce delay
(see also Burrows, Dessart, \& Livne 2007). Such neutrino-driven winds are generic after explosion (Burrows, 
Hayes, \& Fryxell 1995), but are too weak to generate a $10^{51}$-erg explosion.  
This progenitor can explode in 1D because its envelope is extremely rarified, 
and the wind emerges almost immediately, but the associated energy is uncharacteristically 
meager ($\sim$$2.5\times 10^{49}$ ergs, more than an order of magnitude smaller than the canonical value).  
It explodes spherically because there is no inhibiting accretion tamp and can not be generic.  

In the 1980's, Wilson (1985) suggested the ``delayed" mechanism of explosion, wherein
the stalled shock was revived by neutrino heating, but after hundreds of milliseconds.
This mechanism was championed by Bethe \& Wilson (1985), who introduced the concept 
of the ``gain region" behind the shock, where net neutrino energy deposition was positive. 
Though his calculations were done in 1D, Wilson required ``neutron-finger" convection below the 
neutrinospheres to boost the driving neutrino luminosities.  Such instabilities, and the corresponding 
luminosity boosts, have been called into question by Bruenn \& Dineva (1996) and Dessart et al. (2006a).

In the 1990's, the available computer power enabled detailed numerical explorations 
beyond spherical symmetry into two dimensions (Burrows and Fryxell 1992; Herant, Benz,
and Colgate 1992; Herant et al. 1994; Burrows, Hayes, and Fryxell 1995;
Janka and M\"uller 1996), with some of these 2D calculations leading 
to delayed neutrino-driven explosions.  These were the first simulations of 
collapse and explosion that demonstrated the existence and potential centrality of 
hydrodynamic instabilities and neutrino-driven convection in the supernova 
phenomenon.  Fryer and Warren (2002,2004) and Fryer \& Young (2007) later performed 3D simulations, but used 
an SPH hydro code that may not have adequately handled the resultant instabilities 
and turbulence. Nevertheless, those were some of the first 3D simulations in supernova 
theory that attempted to include the relevant physics. 

Mezzacappa et al. (1998) challenged the notion that 
multi-D effects were central to the supernova phenomenon, suggesting 
that precision spherical simulations for which detailed neutrino 
transport was incorporated would be necessary to demonstrate robust 
explosions.  However, this is not the current view, with multi-D effects first explored in the 1990's now assuming 
center stage. More recently, using purely hydrodynamic calculations, Blondin, Mezzacappa, \& DeMarino (2003) 
showed that the shock itself, even without neutrino heating and the consequent convection,
could be unstable, introducing the ``standing-accretion-shock instability" (SASI).
This hydrodynamic phenomenon has been studied analytically and in detail by Foglizzo et al. (2006,2007).
One can not easily disentangle neutrino-driven convection and the SASI, and it may be that
neutrino-driven convection is predominant (Fernandez \& Thompson 2009).  The major contribution of the
Blondin et al. paper was to highlight the need to perform 2D calculations
over the full 180$^{\circ}$ so as not to suppress the interesting dipolar component.
Most previous calculations had been performed on a 90$^{\circ}$ wedge.
Otherwise, the ``SASI" had been manifest and observed naturally in previous multi-D simulations.

Buras et al. (2006ab), using their code MuDBaTH, obtained an 
explosion in 2D with the 11.2-M$_{\odot}$ progenitor of Woosley \& Heger (2006).  
However, this model's explosion was underpowered by an order of magnitude.  
More recently, Marek \& Janka (2009) obtained an explosion of a
rotating 15-M$_{\odot}$ progenitor (their model M15LS-rot), a more representative star.  However, they
experienced a long delay to explosion of more than 600 milliseconds,
by which time the mass in the gain region was too low to absorb enough neutrino
energy to achieve more than $\sim$$2.5\times 10^{49}$ ergs by the end of their simulation.  Moreover,
the shock wave was not followed beyond $\sim$600 km and they used a soft nuclear
equation of state with an incompressibility at nuclear densities, $K$, of 180 MeV. 
Their model with $K$ = 263 MeV did not explode. An incompressibility nearer 
240$\pm$20 MeV is currently preferred by measurement (Shlomo, Kolomietz, \& Col\`{o} 2006).
In addition, Marek and Janka's high-resolution run of the same model (M15LS-rot-hr) did not explode, 
though it was continued to nearly the same final post-bounce epoch as their 
exploding model M15LS-rot.

The Oak Ridge supernova group, led by A. Mezzacappa and S. Bruenn, is paralleling in many ways the
work of the Marek \& Janka (2009), but is using its CHIMERA code (Bruenn et al. 
2007,2010; Mezzacappa et al. 2007) and Bruenn's 1D multi-group, flux-limited 
diffusion algorithm (Bruenn 1985) in the multiple 1D ``ray-by-ray" transport approximation
\footnote{Also used by the Marek \& Janka (2009), and first introduced into supernova 
theory in more primitive form by Burrows, Hayes, \& Fryxell (1995).}. 
As do Marek \& Janka (2009), the ORNL team employs the lower-$K$, softer nuclear EOS.  They have found robust explosions 
for a variety of progenitors (Bruenn et al. 2007,2010;  Mezzacappa et al. 2007) and have suggested 
that the mechanism is a combination of neutrino heating, nuclear burning (of infalling 
oxygen at the shock), the SASI, and inelastic neutrino-electron 
and neutrino-nucleon scattering.  However, the latter should be a subdominant 5$-$10\% effect.  
The potential role of nuclear burning on infall had been studied and eliminated 
earlier (Burrows \& Lattimer 1985; H.-T. Janka, private communication) and though included in many other 
previous simulations has not been seen to have played a positive role. Infalling oxygen should burn
long before reaching the shock wave that has stalled near $\sim$150$-$200 kilometers
and, moreover, the amount of nuclear energy available is rather small. 
Why the results of these two groups differ so substantially
(when their approaches are ostensibly so similar), in both explosion energy when 
explosions are claimed, and in whether they see explosions at all, is a puzzle.

Those using the fully 2D radiation/hydrodynamics code VULCAN (Livne 1993; Livne et al. 2004,2007; 
Burrows et al 2006, 2007ab), apart from obtaining neutrino-driven explosions 
in the contexts of a 8.8-M$_{\odot}$-progenitor (Burrows, Dessart, \& Livne 2007)
and accretion-induced-collapse (Dessart et al. 2006b), do not witness neutrino-driven 
explosions for most progenitors.  VULCAN/2D incorporates an 
implicit, time-dependent, multi-energy-group transport scheme 
that has both multi-angle (S$_n$) and flux-limited (MGFLD) variants in two spatial 
dimensions.  VULCAN is the first and (to date) only multi-group supernova code to operate in     
2D in both the hydro and radiation sectors.  Interestingly, using a multi-angle variant of VULCAN, Ott et al. (2008)
have shown that for non-rotating models the results for the multi-angle and MGFLD simulations are 
similar. VULCAN also incorporates ``2.5-D" MHD and, hence, is the only code currently employed in supernova studies 
with both multi-group transport and MHD capabilities.  With it, Burrows et al. (2007c) have explored
MHD-driven jet explosions (LeBlanc \& Wilson 1970; Bisnovatyi-Kogan, Popov, 
\& Samokhin 1976; Akiyama \& Wheeler 2005; Wheeler \& Akiyama 2007), but concluded that 
such explosions, which require very rapid rotation at bounce that is very unlikely
to be generic, might obtain only in the rare ``hypernova" case and could not explain the typical supernova.

However, using VULCAN and after simulating for around a second after bounce, Burrows 
et al. (2006, 2007a) observed vigorous dipolar g-mode oscillations
that damp by the asymmetric emission of sound waves that steepen into shock waves.
This acoustic power has been adequate to explode all progenitors, but
would be aborted if the neutrino mechanism, or some other mechanism, obtained
earlier.  This acoustic mechanism is controverial, takes a long time (many seconds) to achieve
explosion energies of $\sim$10$^{51}$ ergs, and has not been seen by other groups.
However, no other group has calculated for the physical time necessary 
to witness vigorous g-mode excitation.  Unlike all VULCAN simulations, most 
other simulations have been done with the inner core in 1D\footnote{The VULCAN modelers used an enabling 
Cartesian grid in the inner 30$-$50 kilometers.}.  As a result, we suggest that 
other groups have not simulated long enough to see this phenomenon. Moreover, we suspect that
installing a spherical inner core will partially suppress the excitation of a core g-mode 
that is fundamentally dipolar, not spherical, and whose amplitude should be largest 
at the center.  A solid criticism of the acoustic mechanism comes from Weinberg 
\& Quataert (2008), who using approximate analytic techniques, suggest that the non-linear 
excitation of dissipative daughter modes via a parametric resonance
might render the amplitudes of these dipolar g-mode oscillations too small to power
a supernova. Unfortunately, the wavelengths of these daughter modes are smaller
than reasonable grid spacings and can not easily be captured by extant supernova codes.
Hence, whether a very delayed acoustic power mechanism for explosion is at all viable 
in any circumstance remains to be seen. 

However, the delayed neutrino mechanism seems compelling, if only it can be shown 
to work robustly and to give canonical explosion energies of $\sim$10$^{51}$ ergs generically.  
Burrows \& Goshy (1993) suggested that the neutrino-driven explosion is a critical phenomenon.
There is a critical luminosity at a given mass accretion rate into the shock above which
there is no accretion solution and the mantle explodes.
Recently, Murphy \& Burrows (2008) have shown that
this critical luminosity decreases and the neutrino-matter coupling
efficiency increases by $\sim$30\% in going from 1D to 2D.
Murphy \& Burrows (2008) have diagnosed the increase in heating efficiency in 2D (vis \`a vis 1D) 
as the increase in the average dwell time (Thompson, Quataert, \& Burrows 2005) 
of matter in the gain region due to neutrino-driven convection and the SASI, 
before it settles into the cooling region and onto the inner core. 
What hasn't been shown, and is our goal with this paper, is the corresponding effect in going from 2D to 3D.

The absence in VULCAN of inelastic scattering terms 
and any corrections for general relativistic effects might compromise its 
conclusions and be the reason those using VULCAN are not seeing neutrino-driven explosions in the generic
case, or when other groups obtain such explosions (however tepid). However, the inelastic terms are 
small, and relativity's effects nearly cancel.  A more plausible explanation for the 
ambiguity rife in the field and the marginality (and rarity) of neutrino-driven explosions  
among published simulations is that the neutrino mechanism may not work well in 2D. It may be that the 
neutrino mechanism is truly viable only in 3D and that 3D effects have been the missing 
ingredient needed to explain supernova explosions with delayed neutrino driving.  This is the thesis of our paper.
We posit that the viability of the neutrino-driven mechanism for core-collapse supernova explosions
and the energy of explosion (when they explode) are monotonic functions of dimension.  We note that 
it is only recently, with the advent of computers of sufficent size 
and speed, that theorists have been able to test such an hypothesis. 
Furthermore, we suggest that the ambiguity of the extant 2D results is a symptom of its marginality
in 2D.  Moreover, as did Murphy \& Burrows (2008), we conjecture that the increased dwell time in the gain region in going to 3D,
due both to the extra degree of freedom for non-radial motion and the character of the cascade in 3D turbulence 
to small turbule sizes\footnote{In 2D, the turbulent cascade is artifically inverted and flows to large scales.},
renders the heating efficiency and all the other ancillary quantities that support
explosion adequate to transform ``marginality and ambiguity" into ``robustness and 
viability" (\S\ref{results}).

\section{CASTRO Code and General Methodology}
\label{castro}

To address these issues in a full 3D context, we have employed
CASTRO (Compressible ASTRO; Almgren et al. 2010), a new multi-dimensional
radiation/hydrodynamic code. CASTRO is an Eulerian, structured grid, compressible,
radiation/hydrodynamics code that incorporates adaptive mesh refinement
(AMR).  It uses a second-order unsplit piecewise-parabolic
(PPM) Godunov methodology for general convex equations of state.
The most general treatment of self-gravity in CASTRO uses multigrid to
solve the Poisson equation for the gravitational potential.  For
the calculations presented here, the monopole approximation is used.

AMR in CASTRO uses a nested hierarchy of rectangular
grids that are simultaneously refined in both space and time.
CASTRO uses a recursive integration procedure in which coarse grids are
advanced in time, fine grids are advanced multiple steps to reach the
same time as the coarse grids, and the data at different levels are
then synchronized.  For regridding, an error estimation procedure
evaluates where additional refinement is
needed and grid generation procedures dynamically create or destroy
fine grid patches to achieve the desired local resolution.
CASTRO is implemented within the BoxLib framework that handles data
distribution, communication, memory management, and I/O
for parallel architectures.

To most closely parallel the approach of Murphy \& Burrows (2008),
while generalizing to three dimensions, our calculations were
done with CASTRO using the simple neutrino heating and cooling algorithm
described in Murphy \& Burrows (2008).  
We solve the fully compressible equations of hydrodynamics:
\begin{eqnarray}
\partial_t\rho &=& -\nabla\cdot\left(\rho \mathbf{u}\right) \label{eq:hydro}\\
\partial_t\left(\rho\mathbf{u}\right) &=& -\nabla\cdot\left(\rho \mathbf{u}\mathbf{u}\right) 
- \nabla p + \rho \mathbf{g}\nonumber\\
\partial_t\left(\rho E\right) &=&
-\nabla\cdot\left(\rho\mathbf{u}E\right)+p\mathbf{u}+
\rho\mathbf{u}\cdot\mathbf{g}+\rho(\mathcal{H}-\mathcal{C})\, , \nonumber
\end{eqnarray}
where $\rho$, $T$, $p$, $\mathbf{g}$, and $\mathbf{u}$ are the fluid density, temperature, pressure, 
gravitational acceleration, and velocity.  The total energy is given by $E = e + \frac{1}{2}u^2$, where $e$ 
represents the internal energy.  The equation of state provides closure and 
is a function of $\rho$, $T$, and electron fraction, Y$_{e}$. While a sophisticated nuclear
equation of state is used (Shen et al. 1998), the neutrino heating and
cooling rates behind the stalled shock wave via the
super-allowed charged-current reactions involving free nucleons assume
a given electron neutrino and anti-electron neutrino luminosity (taken
to be the same).  This luminosity is varied from simulation to
simulation, but is held constant during a simulation. The neutrino 
heating, $\mathcal{H}$, and cooling, $\mathcal{C}$, rates are those derived in Janka 2001,
used in Murphy \& Burrows (2008), and given by
\begin{eqnarray}
\mathcal{H} &=& 1.544\times 10^{20} \left(\frac{L_{\nu_e}}{10^{52}\ 
{\rm erg}\ {\rm s}^{-1}}\right) \left(\frac{T_{\nu_e}}{4\ {\rm MeV}}\right)^2\times\label{heating}\\
&& \left(\frac{100 {\rm km}}{r}\right)^2 \left(Y_{\rm n} + Y_{\rm p}\right) 
e^{-\tau_{\nu_e}} \left[\frac{{\rm erg}}{{\rm g}\ {\rm s}}\right]\nonumber
\end{eqnarray}

and

\begin{equation}
\mathcal{C} = 1.399\times 10^{20} \left(\frac{T}{2\ {\rm MeV}}\right)^6 
\left(Y_{\rm n} + Y_{\rm p}\right) e^{-\tau_{\nu_e}}
\left[\frac{{\rm erg}}{{\rm g}\ {\rm s}}\right]\label{cooling}\, ,
\end{equation}
where $L_{\nu_e}$ is the electron neutrino driving luminosity, $T_{\nu_e}$ 
is the electron neutrino temperature, $r$ is the distance from the center of 
the star, $Y_{\rm n}$ and $Y_{\rm p}$ are the neutron and proton fractions 
and $\tau_{\nu_e}$ is the electron neutrino optical depth.  Note that these 
approximations assume that the $L_{\nu_e}=L_{\bar{\nu}_e}$.   

This approach, using eqns. (\ref{heating}) \& (\ref{cooling}) 
in place of full transport, enabled the extensive (and computationally 
expensive) parameter study in 1D, 2D, and 3D we report here.  The 
single progenitor we employ for all our runs is the non-rotating,
solar-metallicity 15-M$_{\odot}$ red-supergiant model of Woosley \& Weaver (1995),
but we need only the inner $\sim$10,000 km for our simulations of core collapse. 
The calculations commence at the onset of infall and are
carried out to and after bounce. The explosion time is formally 
determined when the average shock radius reaches 400 km and 
does not recede during subsequent evolution (see Table 1). Many of the 1D and 2D simulations
are carried to beyond a second after bounce (up to $\sim$1.4 seconds).
The 3D simulations were all carried out to beyond 400 milliseconds
after bounce. The Liebend{\"o}rfer et al. (2005) scheme
for determining the electron fraction on infall was used, which
involves tying Y$_e$ to the mass density achieved for densities
below the trapping density.

Interior to a radius of 200 km, all of our simulations have zones 
smaller than a kilometer. Exterior to 200 km, the resolution dynamically 
follows the shock via adaptive grids.  In 3D, our Cartesian domain
is a cube of length 10,000 km. The domain is discretized at the coarsest
level with a $304^3$ uniform grid, and several levels of adaptive refinement
are used so that cells at the finest level are a factor of 64 finer
in each direction. This results in a resolution of $\sim$0.5 km (which we
refer to as the ``effective" resolution) in the interior 200 km of the
star and in other regions at the highest refinement level. For our
axisymmetric simulations, we employ a 2D domain that is 10,000 km
by 5,000 km, discretized with a uniform coarse grid of 256 by 128 cells.
All of our 2D simulations cover the full 180$^{\circ}$ angular range from north 
to south pole. Employing the same refinement criteria used in 3D yields an effective
resolution of $\sim$0.6 km for our 2D simulations.  For our
spherically symmetric (1D) simulations, we can obtain, and have
tested, higher resolution runs.  Additional resolution leads to
minimal differences ($\lesssim$10 ms) in the reported explosion times.
The 1D results presented in this paper are for a radial domain of 256 coarse
cells (with similar refinement to that employed in the multi-D simulations),
distributed over 5,000 km for an effective resolution of $\sim$0.3 km.

\section{Results}
\label{results}

\begin{figure}
\plotone{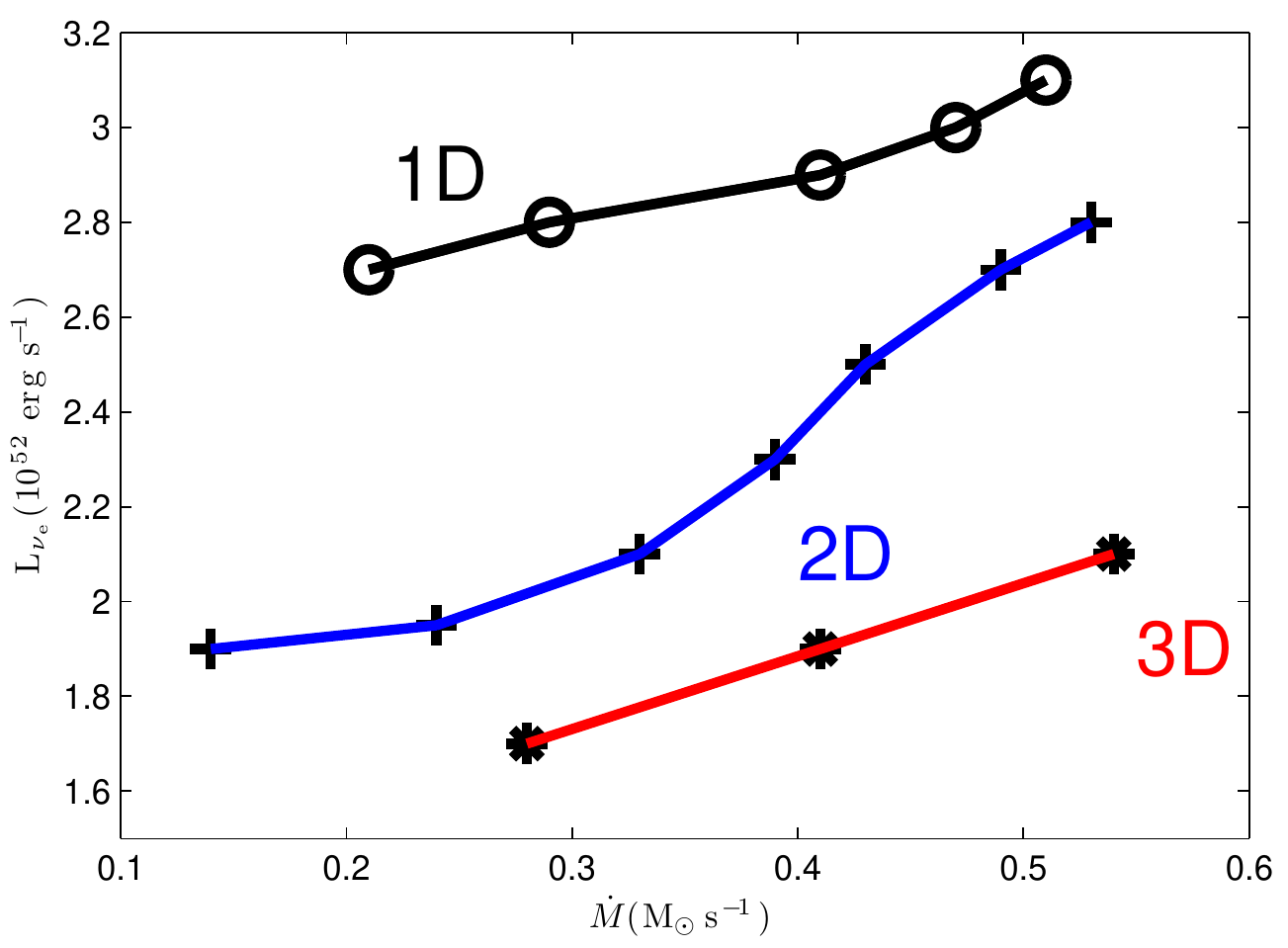}
\caption{Critical curves in electron-neutrino driving luminosity ($L_{\nu_e}$) 
versus accretion rate ($\dot{M}$) for calculations performed in 1D (black), 2D (blue), and 3D (red). 
The luminosity is in units of $10^{52}$ ergs s$^{-1}$ and the mass 
accretion rate ($\dot{M}$) is in solar masses per second. 
Importantly, we include results in 3D. Note that the driving electron neutrino 
luminosity is always accompanied by an associated anti-electron neutrino luminosity (L$_{\bar{\nu}_e}$).
See \S\ref{results} for a discussion of the meaning of this plot and its salient features.} 
\label{fig1}
\end{figure}

Depicted in Fig. \ref{fig1} are critical curves in driving luminosity 
versus accretion rate ($\dot{M}$) through the shock for calculations performed in 1D, 2D, and 3D,
all else being equal. The luminosity is in units of $10^{52}$ ergs s$^{-1}$ and the mass
accretion rate is in solar masses per second (M$_{\odot}$ s$^{-1}$).  $\dot{M}$
is $4\pi r^2 \rho |v_R|$ in the infalling material just exterior to the shock wave.  
Above each curve the core explodes and below each curve it does not.  
As shown in Murphy \& Burrows (2008), the 1D curves approximately 
recapitulate the 1D analytic theory found in Burrows \& Goshy (1993).
The important result in this paper is the position of the 3D curve.  It is $\sim$15$-$25\%
below the corresponding 2D curve, which is in turn $\sim$30\% below the 1D curve.
Moreover, the advantage of going to 3D is larger for higher driving lumniosities.  
The upshot is that the magnitude of the driving neutrino luminosity necessary to ``supernova"
a given core for a given mass accretion rate through the shock is reduced in 
going from 1D to 3D by $\sim$40$-$50\%, a rather large, perhaps enabling, effect.  
 
Note that in our study each run is performed for a given neutrino and anti-neutrino 
luminosity, while the accretion rate onto the inner core evolves with time, running
from high to low values of $\dot{M}$ in a way determined by the initial inner density
profile of the chosen progenitor star (in this case the 15-M$_{\odot}$ progenitor 
of Woosley \& Weaver 1995).  Hence, the $\dot{M}$ is not fixed during a run and if another
progenitor model had been chosen the temporal evolution of $\dot{M}$ would have been different.  
When the $\dot{M}$ reaches the value for a given fixed L$_{\nu_e}$ at which the core 
explodes, we identify this L$_{\nu_e}$$-$$\dot{M}$ pair as a point on the critical curve 
(Fig. \ref{fig1}).  Performing this exercise for a range of luminosities maps out the
corresponding critical curve in whatever number of dimensions is being studied. 

Since before explosion the trajectory in L$_{\nu_e}(t)$$-$$\dot{M}(t)$ space
followed by a given progenitor is a weak function of dimension, depending mostly
on the initial progenitor density profile, which in turn determines $\dot{M}$ and thereby
the inner core density, temperature, and Y$_e$ profiles (given an EOS, a neutrino 
transport algorithm, and quasi-hydrostatic equilibrium), changing the dimension of a simulation
is not met with significant compensating shifts in the actual $L_{\nu_e}(t)$ vs. $\dot{M}(t)$ trajectory with which to
compare our critical curves.  The result is an undiminished and monotonic
advantage at higher dimensions of significant magnitude.  The effect of dimension
is seen to be not merely a few percent, but nearly a factor of two in a central 
aspect of the collapse context, the driving neutrino luminosity.  This new result leads us to
suggest that lack of access to 3D computational capabilities has been a major retarding
factor in progress during the last few decades towards the solution to the 
core-collapse supernova problem. 

\begin{figure}
\includegraphics[height=.5\textheight]{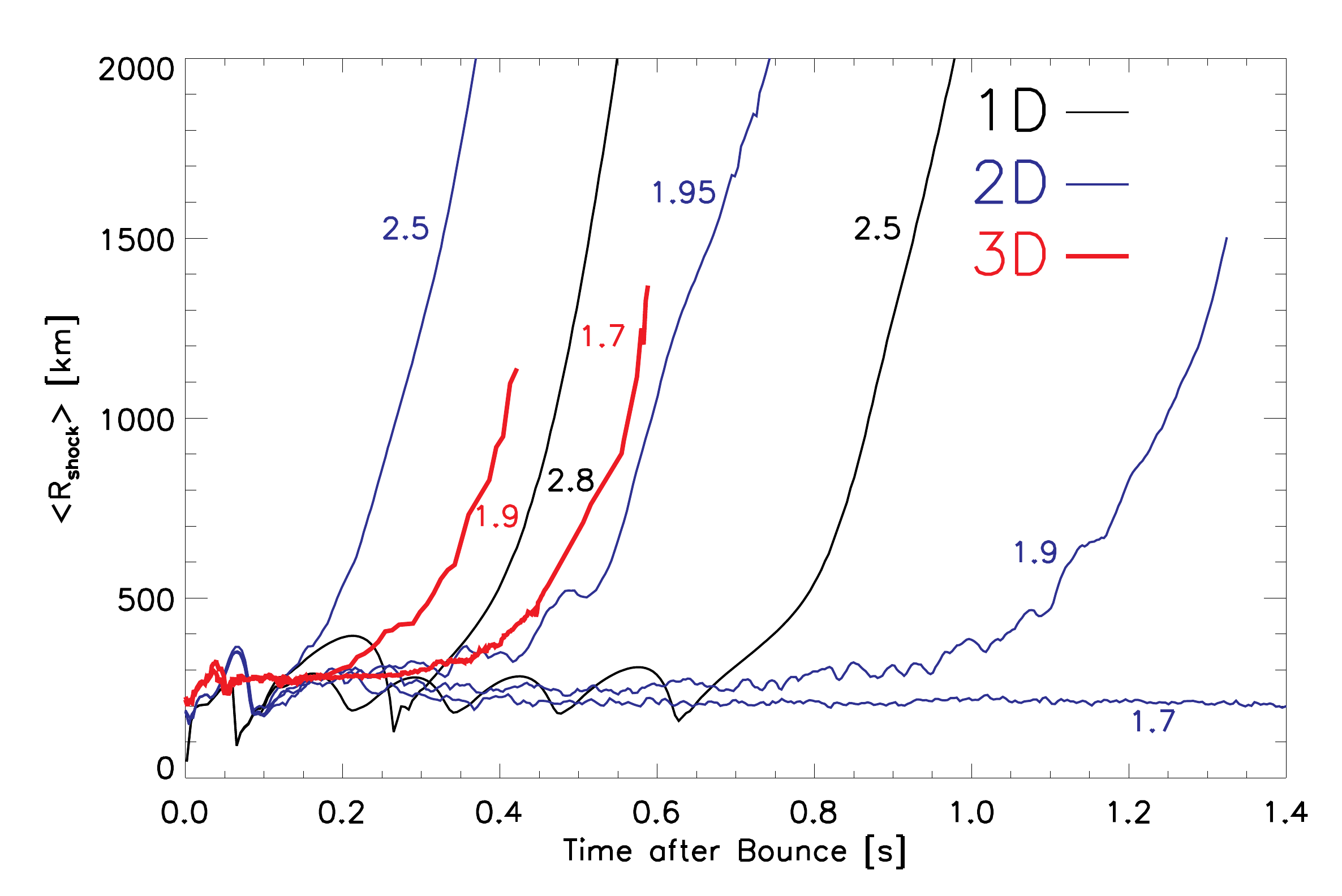}
\caption{Depicted here are curves showing the temporal evolution
of the average radius of the shock (in kilometers) for various driving electron neutrino luminosities and 
for simulations in 1D, 2D, and 3D. The 1D models are rendered as black dashed lines, the 2D models
as solid blue lines, and the 3D models as thick solid red lines.  Each line is indexed
by the corresponding electron-neutrino luminosity of the simulation, in units of $10^{52}$ ergs s$^{-1}$.
Implied is an equal anti-electron-neutrino luminosity. Time is in seconds after bounce.
See the text in \S\ref{results} for a discussion of this figure.}
\label{fig2}
\end{figure}

Plotted in Fig. \ref{fig2} are various curves depicting the temporal evolution
of the average radius of the shock (in kilometers) for various driving electron neutrino luminosities and
for simulations in 1D, 2D, and 3D.  This figure shows that for luminosities for which the 1D and 2D model shocks remain stagnant for
long periods, the 3D models explode much earlier. For instance, at $L_{\nu_e} = 1.9\times 10^{52}$ ergs s$^{-1}$,
in 1D the core doesn't explode even after 1.4 seconds, the 2D core explodes after around $\sim$0.8$-$1.0 seconds,
but the 3D core explodes after only $\sim$250 milliseconds.  In all cases, the time to explosion (if
there is an explosion) is shorter at higher dimension than at lower dimension.
Bruenn et al. (2007,2010) witness explosions in all their recent 2D simulations,
but these results have not been reproduced by others (cf. Marek \& Janka 2009).
Moreover, all their explosions pile up at similar early times, as does the
3D simulation they are currently tending. This suggests that whatever
is causing their explosions does not much distinguish between 2D and 3D in
the way we so clearly see in our suite of simulations.  The reason for this
is unclear, but we note that when we obtain early explosions (in this paper,
at the highest driving neutrino luminosities), the difference in the time
to explosion in 2D and 3D is similarly significantly reduced.  

Table 1 clearly demonstrates that the time to explosion is shorter at higher dimension 
than at lower dimension and provides a more extended compilation of 1D, 2D, and 3D exploding models
and the approximate times at which they explode.  The table is arranged so that overlapping
horizontal rows, though done for a different number of dimensions, have the same driving luminosities. 
This format clearly reveals that, all else being equal, the time to explosion is significantly shorter at higher dimension.  For example,
the 1D model at $L_{\nu_e} = 2.5\times 10^{52}$ ergs s$^{-1}$ explodes around $\sim$0.75 seconds while
the corresponding 2D model explodes near 0.2 seconds.  Interestingly, this is the time at which the 3D model at the much lower
luminosity of $1.9\times 10^{52}$ ergs s$^{-1}$ explodes.  The time it takes the $1.9\times 10^{52}$ ergs s$^{-1}$
model in 3D to lift the average shock radius from $\sim$200-300 kilometers to $\sim$1200 kilometers during the early
explosion phase is $\sim$200 milliseconds, at which point the shock is moving at a speed of $\sim$30,000 km s$^{-1}$.
As indicated on the figure, early in the incipient explosion phase the average shock radius gradually, but steadily,
accelerates.

\begin{figure}
\includegraphics[height=.38\textheight]{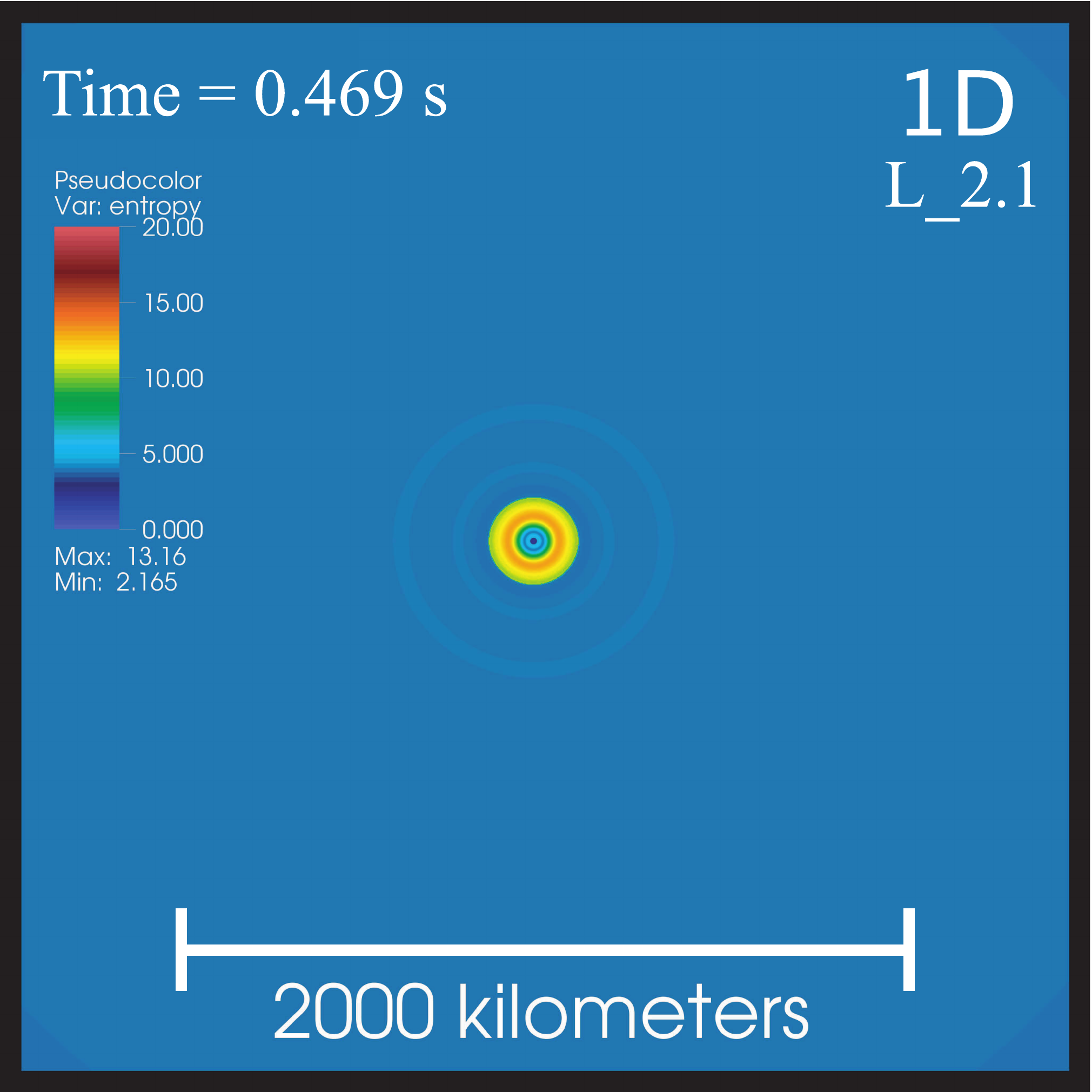}
\includegraphics[height=.516\textheight]{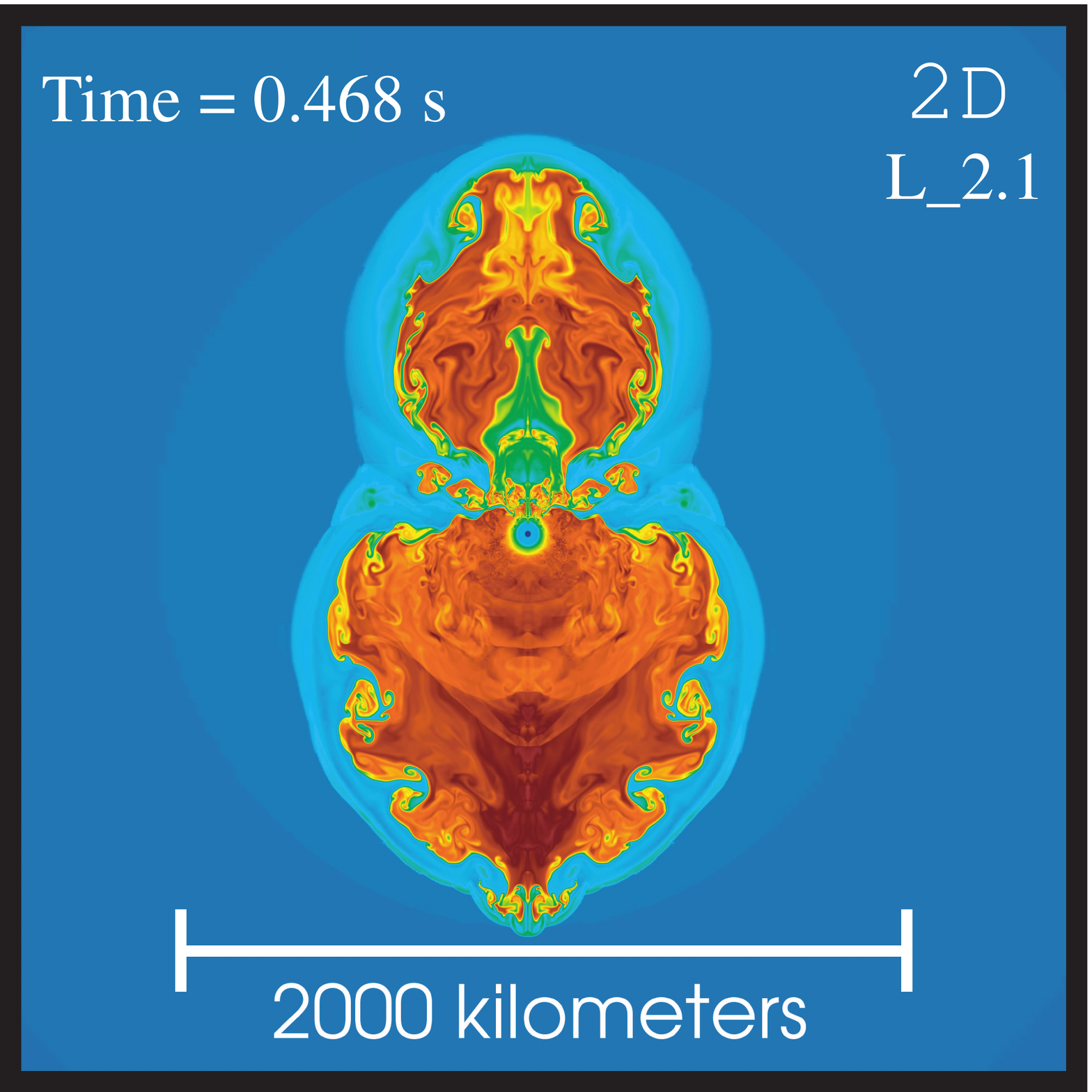}
\includegraphics[height=.35\textheight]{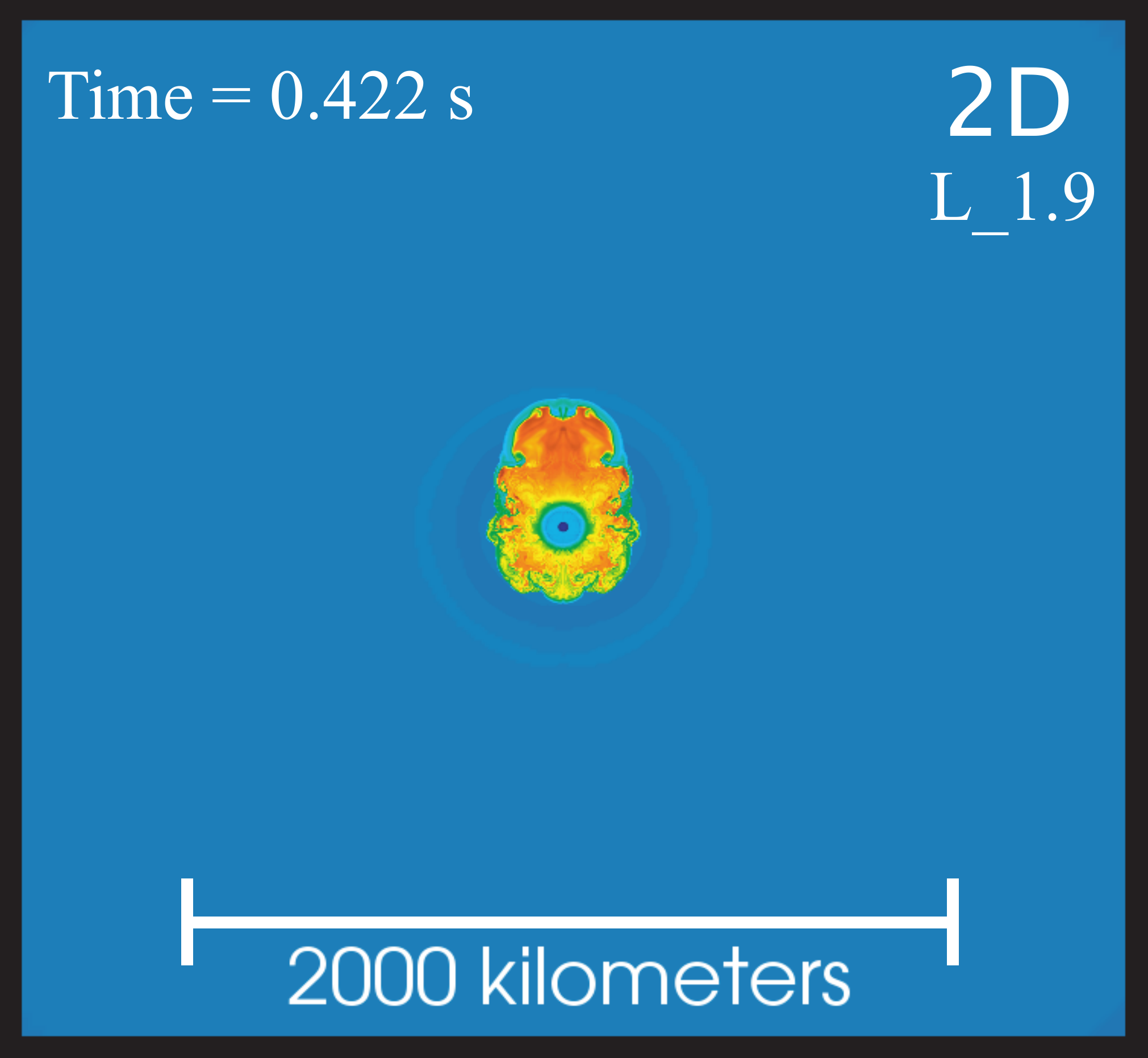}
\includegraphics[height=.35\textheight]{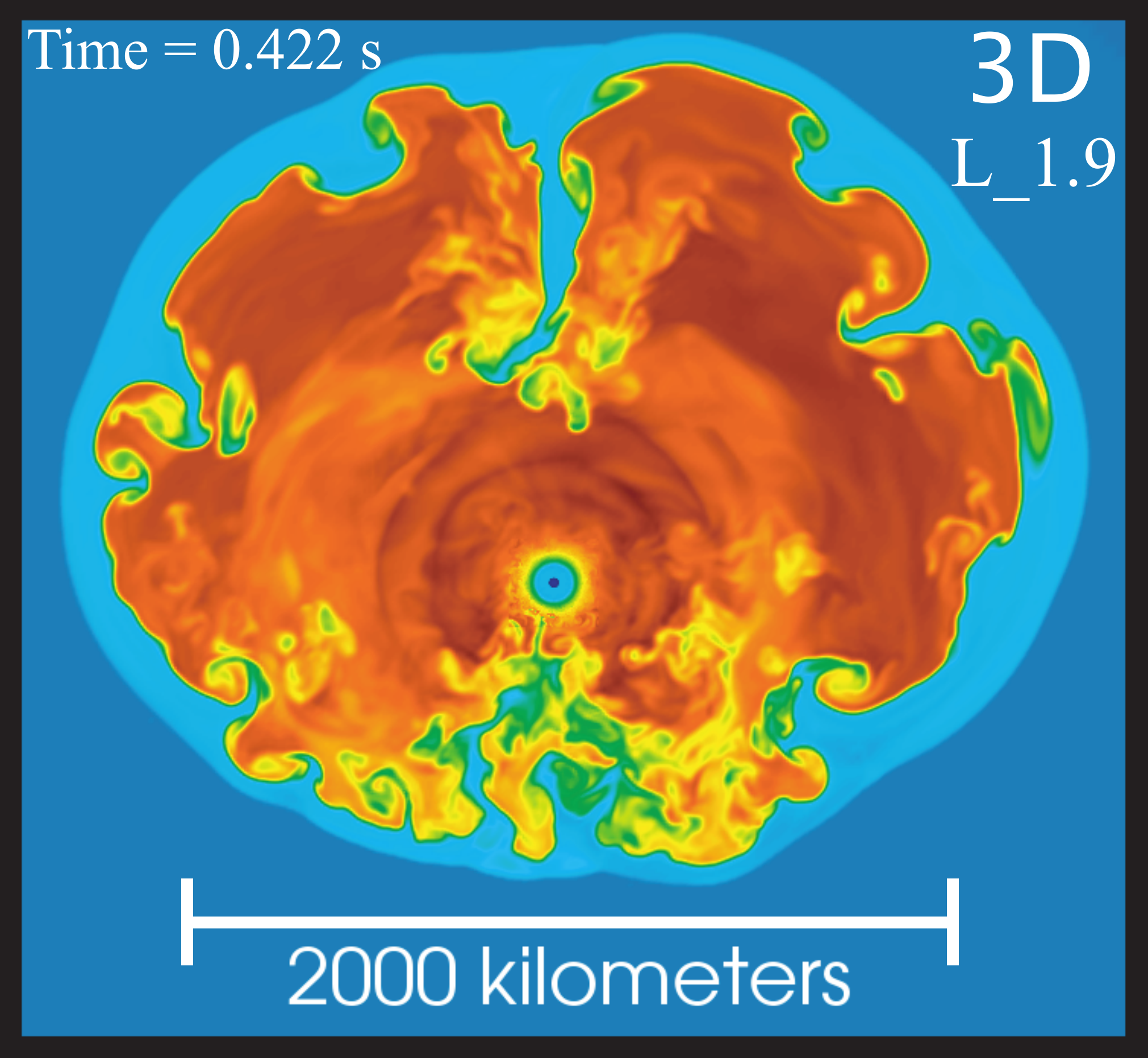}
\caption{In these panels we compare representative entropy maps of two simulations for the same driving neutrino luminosities and
times after bounce, but for different numbers of dimensions.  The top two panels contrast 1D (left) and 2D (right) runs, both 
for an electron-type neutrino luminosity of $2.1\times 10^{52}$ ergs s$^{-1}$ and at $\sim$0.468 milliseconds after bounce.  
The bottom panels compare models in 2D (left) and 3D (right), both for an electron-type neutrino luminosity of $1.9\times 10^{52}$ ergs s$^{-1}$ 
and at 0.422 milliseconds after bounce.  The same colormap is used for all four panels.  Note that
while the two top panels are for the same luminosity and epoch after bounce, only the 2D simulation has
exploded.  Similarly, while the two bottom panels are at the same luminosity and time after bounce, 
the development of the 3D simulation is qualitatively different from that of the corresponding 2D run.
Note also the different general morphologies of the 3D (bottom) and 2D (top) models that explode.  
See the text in \S\ref{results} and \S\ref{conclusions} for a discussion.}
\label{fig3}
\end{figure}

In the panels in Fig. \ref{fig3} we compare representative entropy color maps of simulations for the same neutrino luminosities and
times after bounce, but for different numbers of dimensions.  The top two panels contrast 1D (left) and 2D (right) runs, both
for a luminosity of $2.5\times 10^{52}$ ergs s$^{-1}$ and at $\sim$468 milliseconds after bounce.  Note that
while the 2D run is exploding, the 1D run is not, though the physical model and inputs are otherwise identical.
The positions of the shocks in these simulations, as indicated by the abrupt color transition,
are very different. The bottom panels compare models in 2D (left) and 3D (right), both for a luminosity of $1.9\times 10^{52}$
ergs s$^{-1}$ and at $\sim$422 milliseconds after bounce.  While the shock in the 2D simulation is still stalled, the shock in the 3D
simulation has launched dramatically.  These snapshots together and in conjunction demonstrate pictorially the qualitative
dependence on dimension of the outcome of collapse.  Since supernovae occur in three spatial dimensions, we conclude that this dimensional
dependence is of direct relevance to the viability of the neutrino heating mechanism for core-collapse supernova
explosions.  Moreover, the magnitude of the ``advantage" of going to 3D (see Fig. \ref{fig1}) should dwarf
that of any refinements in the microphysics, such as inelastic neutrino-matter scattering, nuclear burning
upon infall, or general relativity.
The latter are at most ``$\sim$10\%" effects and are small compared with the $\sim$40-50\% effects      
we identify here that are associated with dimension, in particular in going to 3D.

\begin{figure}
\includegraphics[height=.34\textheight]{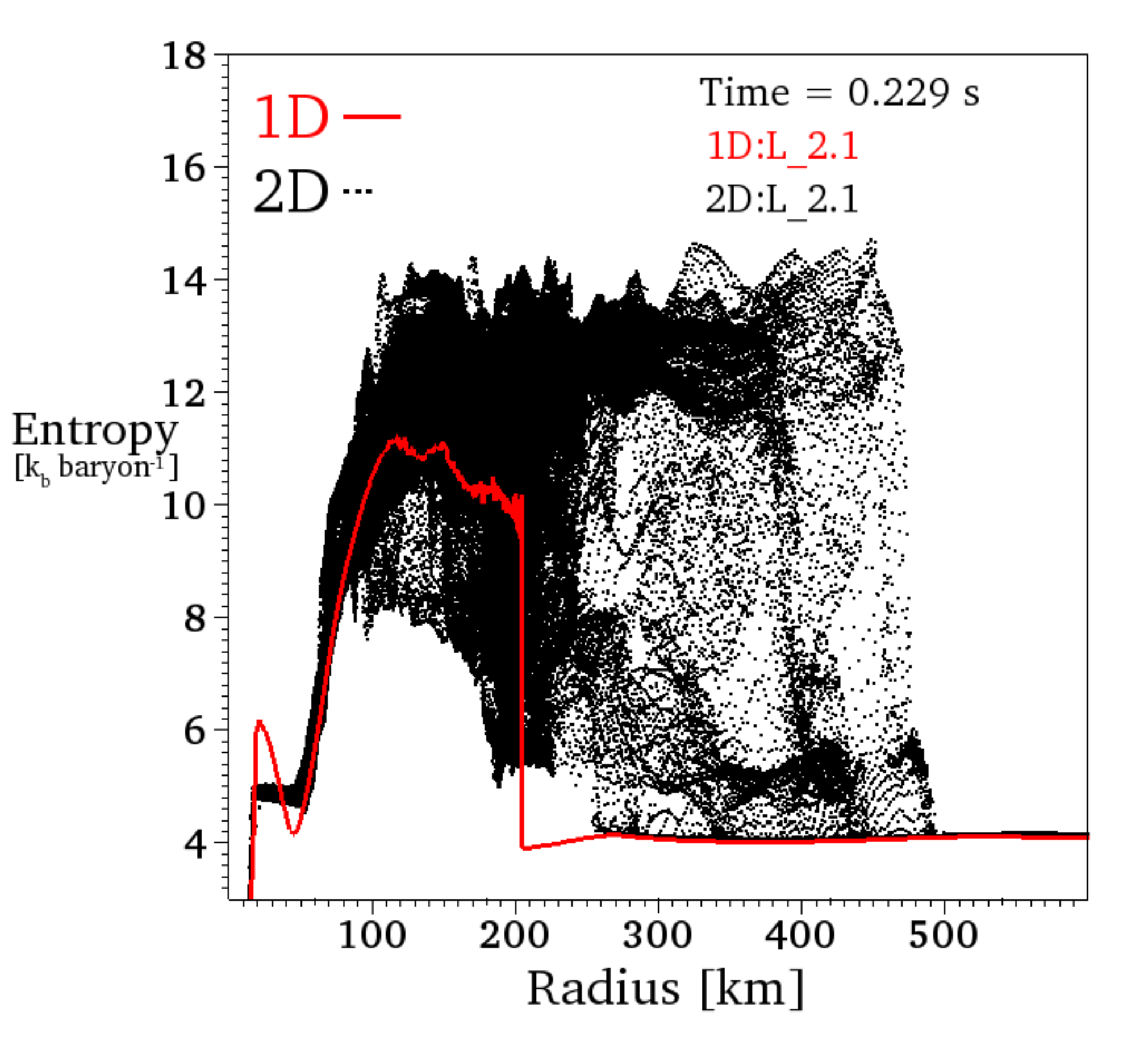}
\includegraphics[height=.34\textheight]{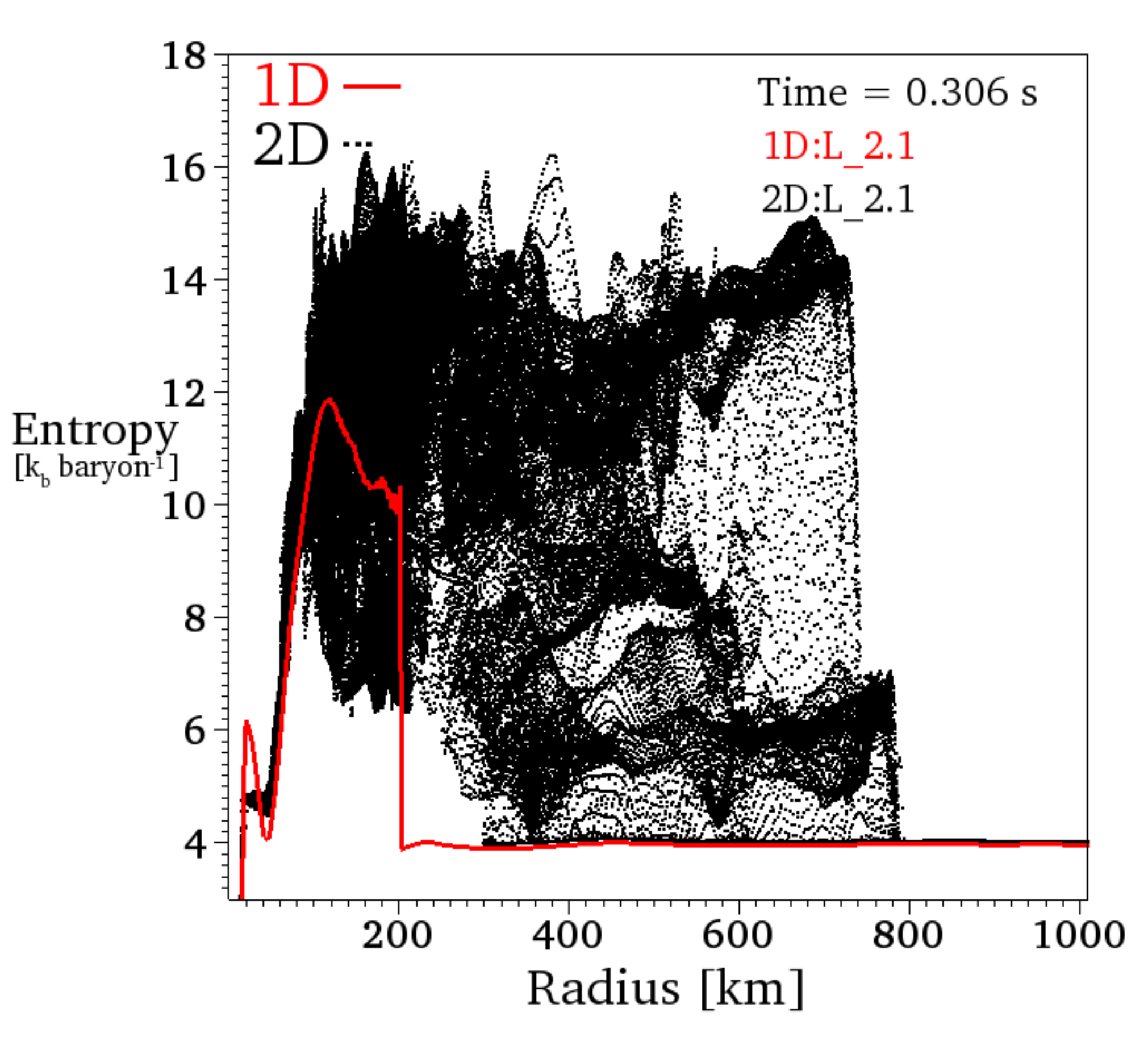}
\includegraphics[height=.34\textheight]{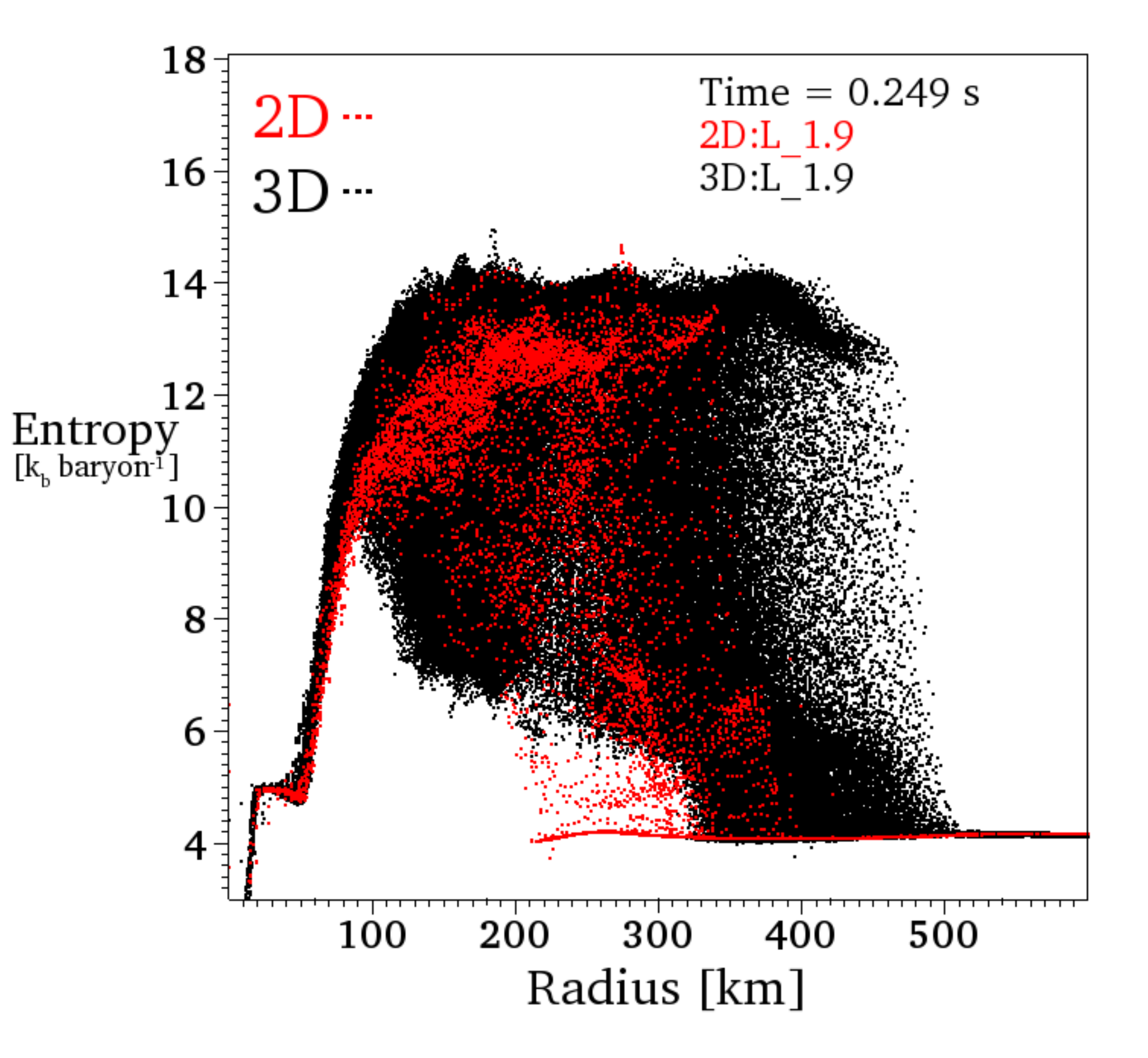}
\includegraphics[height=.34\textheight]{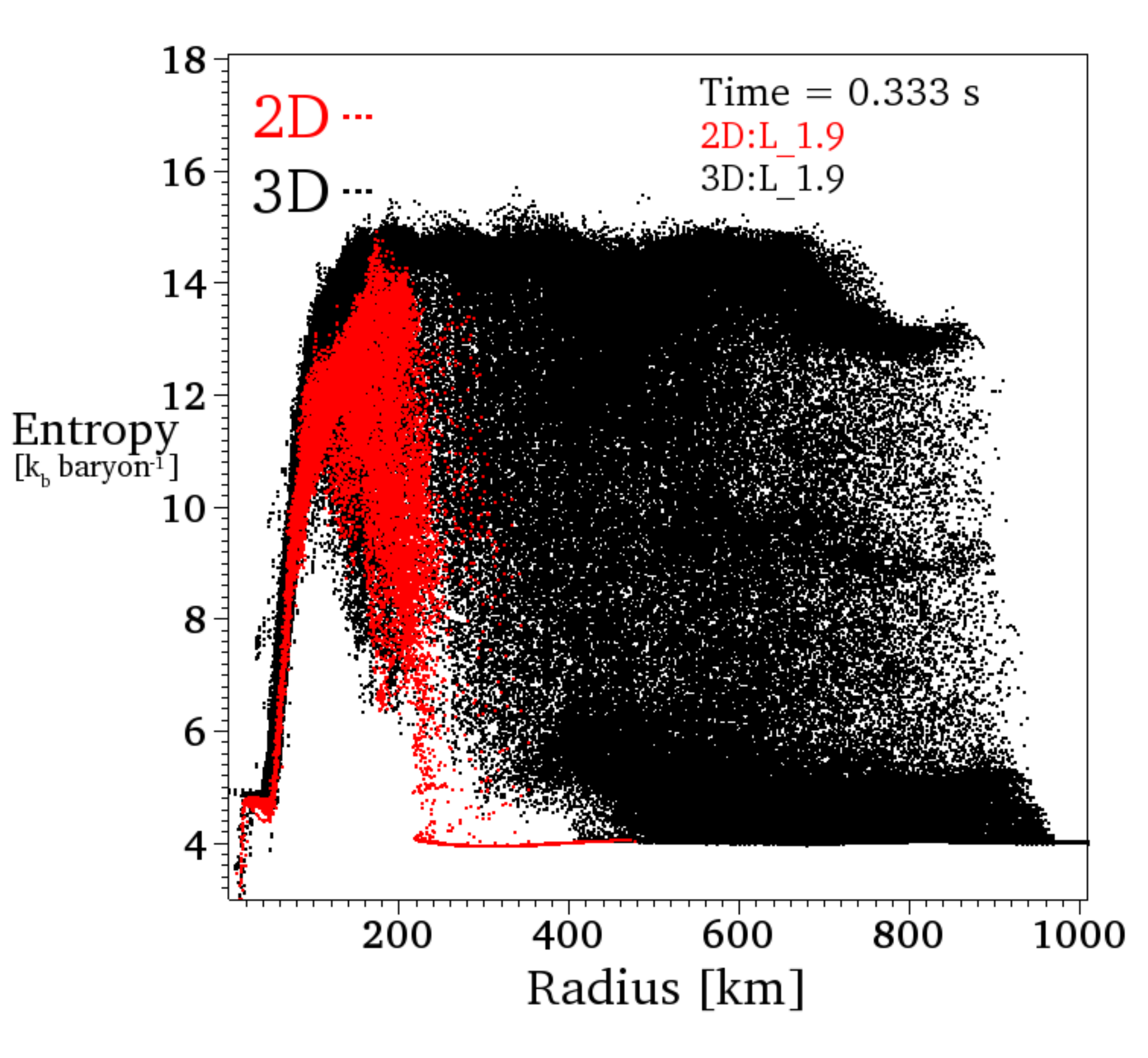}
\caption{Scatter-plot entropy profiles versus radius for two different 
driving luminosities, comparing 1D and 2D (top) and 2D and 3D (bottom).  The 1D(red)-2D(black) comparison
at the top is for L$_{\nu_e}$ = $2.1 \times 10^{52}$ ergs s$^{-1}$ and the 
2D(red)-3D(black) comparison at the bottom is for L$_{\nu_e}$ = $1.9 \times 10^{52}$ ergs s$^{-1}$.
For both top and bottom two different times after bounce are given $-$ the times for both left panels 
are near the onset of the explosion of the associated higher-dimensional model. Note that the  
peak entropies and average entropies are always higher for the higher-dimensional runs,
and that for a given dimension the higher the driving luminosity the higher the 
peak entropy achieved.  Note also that different horizontal radius scales 
are employed for the left and the right panels and that the associated 
model names are given on the plots. See the discussion in \S\ref{results}.}
\label{fig4}
\end{figure}

\begin{figure}
\includegraphics[height=.50\textheight]{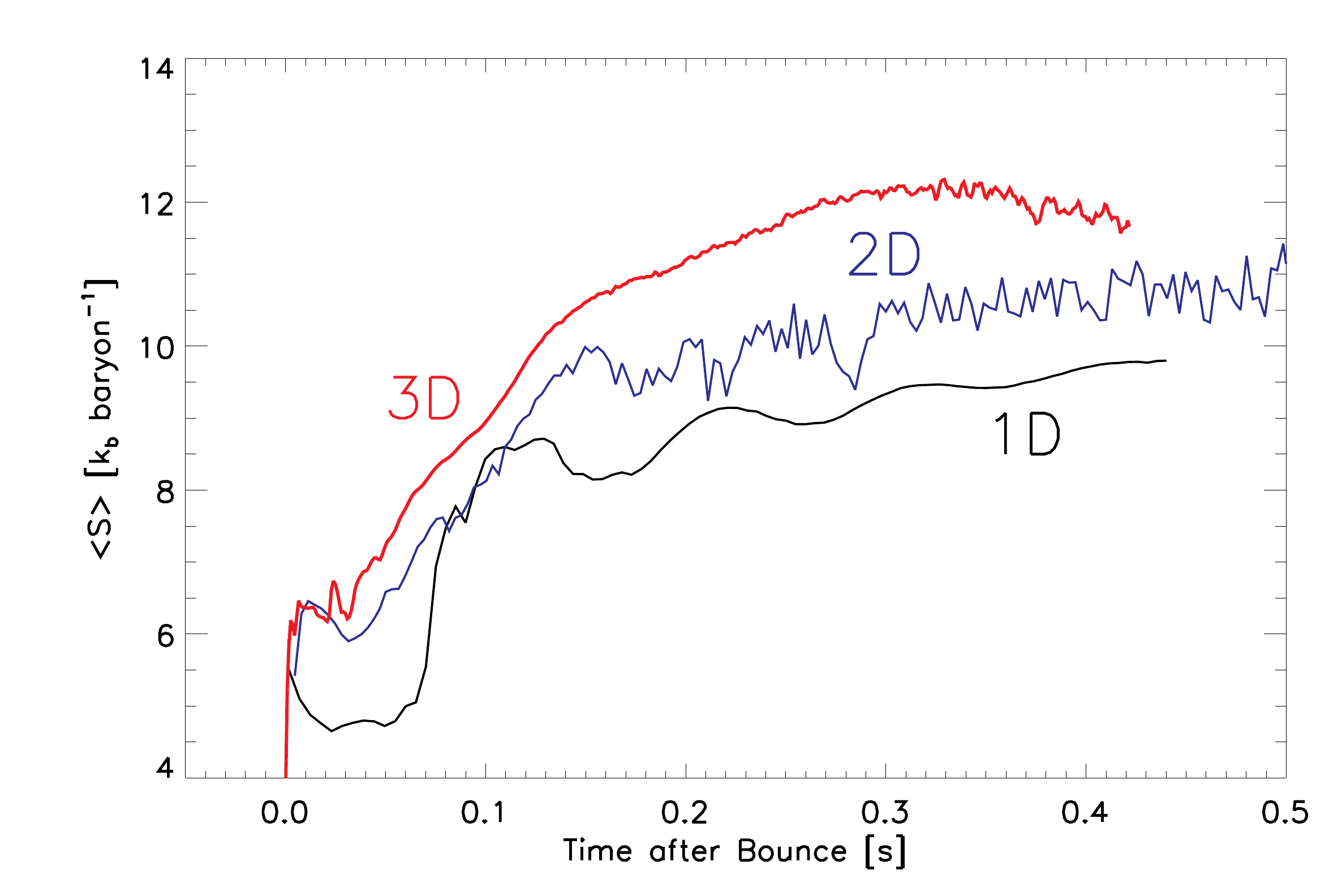}
\caption{The mass-weighted average entropy in the gain region versus 
time after bounce (in seconds) for the 1D(black), 2D(blue), and 3D(red) 
models with $L_{\nu_e}$ = $1.9 \times 10^{52}$ ergs s$^{-1}$. This figure, representative of corresponding 
figures at other driving luminosities, demonstrates the higher entropies
achieved in the gain region behind the shock in going to 3D.  After $\sim$0.2 
seconds after bounce the difference between the 2D and 3D model-average entropies 
is $\sim$1.5 units. See \S\ref{results} for a discussion of the meaning 
and relevance of this figure.}
\label{fig5}
\end{figure}

Figure \ref{fig4} depicts entropy scatter plots near and after the time of explosion
of the corresponding higher-dimensional run for 1) 1D and 2D models (top) and 2) 2D and 3D models (bottom).
In each case, the lower-dimensional model of a given comparison (either top or bottom) 
has not exploded during the time spanned by the plots.  The top panels are
for L$_{\nu_e} = 2.1\times 10^{52}$ ergs s$^{-1}$ and the bottom panels are for 
L$_{\nu_e} = 1.9\times 10^{52}$ ergs s$^{-1}$.  The times to explosion are given in Table 1.
Figure \ref{fig4} clearly indicates that, all else being equal, higher entropies are achievable 
at higher dimension. Figure \ref{fig5} depicts the corresponding mass-weighted average entropy
in the gain region versus time for the same 1D (black), 2D (blue), and 3D (red) 
models and makes the same point.  Importantly, the average entropy for the 3D run in 
the runup to explosion is $\sim$1.5 units higher than in 2D. 
This expands the physical extent of the gain region, makes it less bound, 
and creates what seems to be a more ``explosive" situation (cf. Figs. \ref{fig1}, 
\ref{fig2}, and \ref{fig3}).  The higher average entropies can be traced to higher average 
dwell times of a Lagrangian parcel of matter in the gain region (Murphy \& Burrows 2008).  
We plan to present a more detailed analysis of this and other effects in a subsequent paper.

\begin{figure}
\centerline{
\includegraphics[height=.5\textheight]{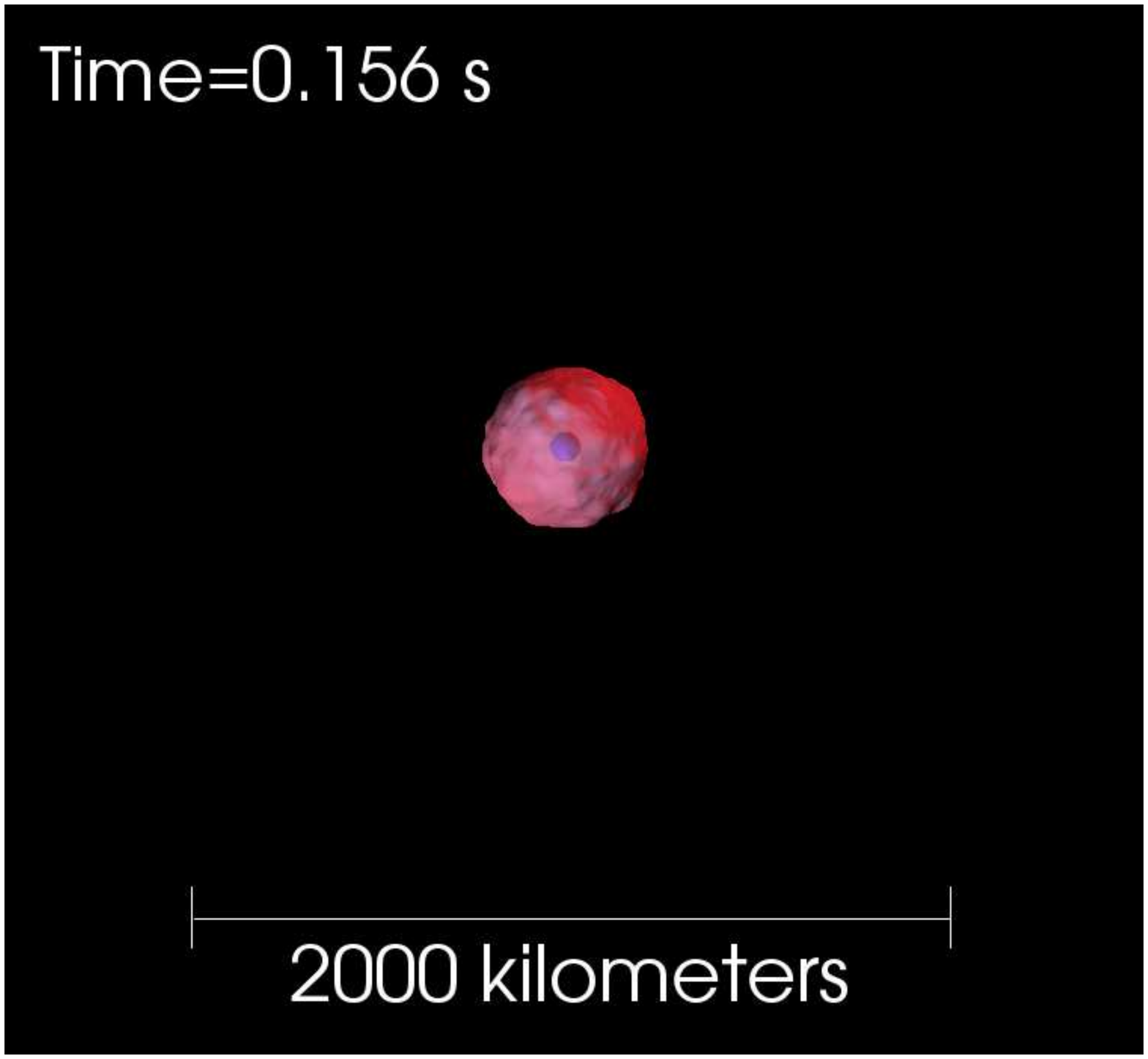}
\includegraphics[height=.5\textheight]{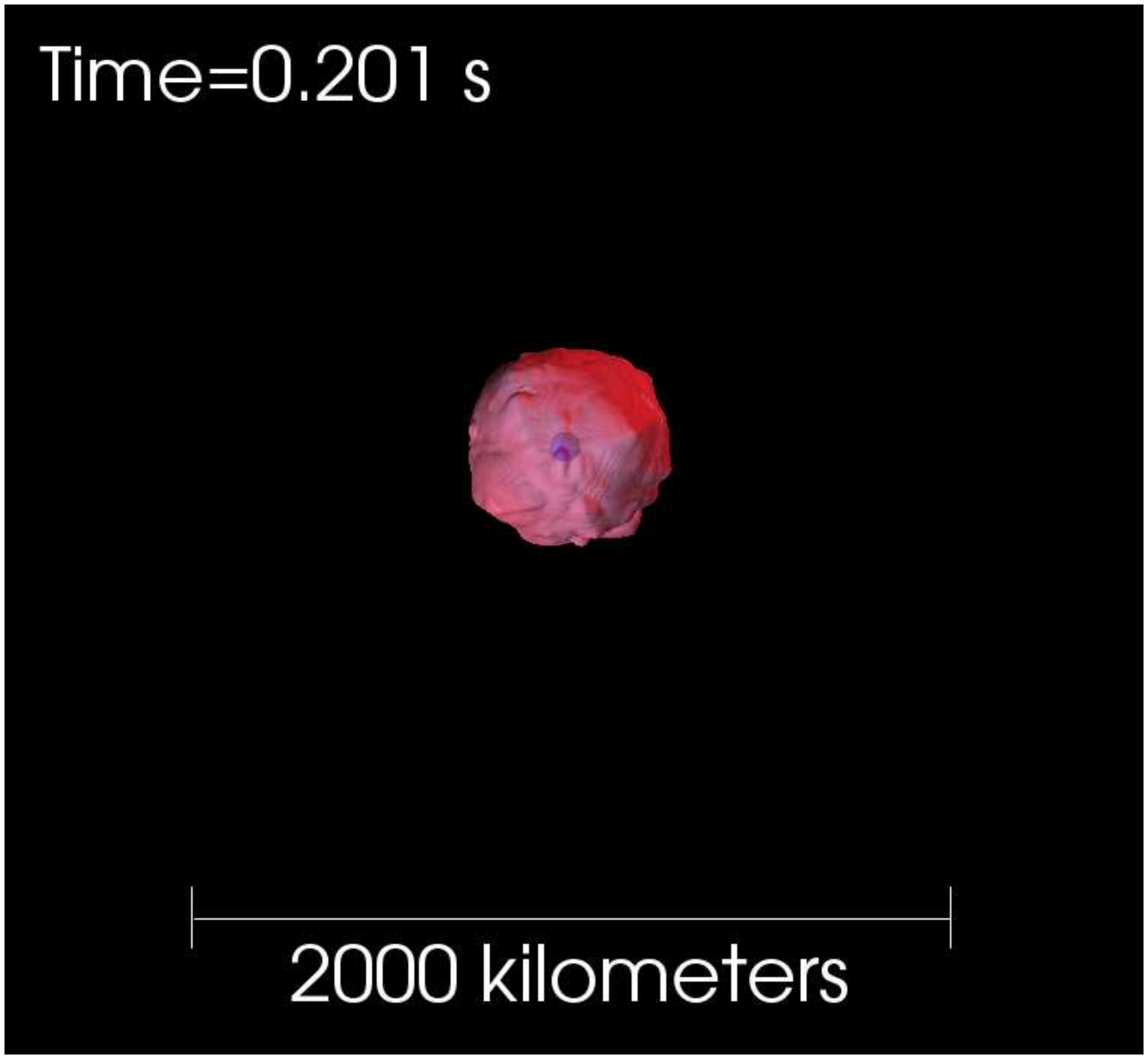}}
\null\vskip-1.5in
\centerline{
\includegraphics[height=.5\textheight]{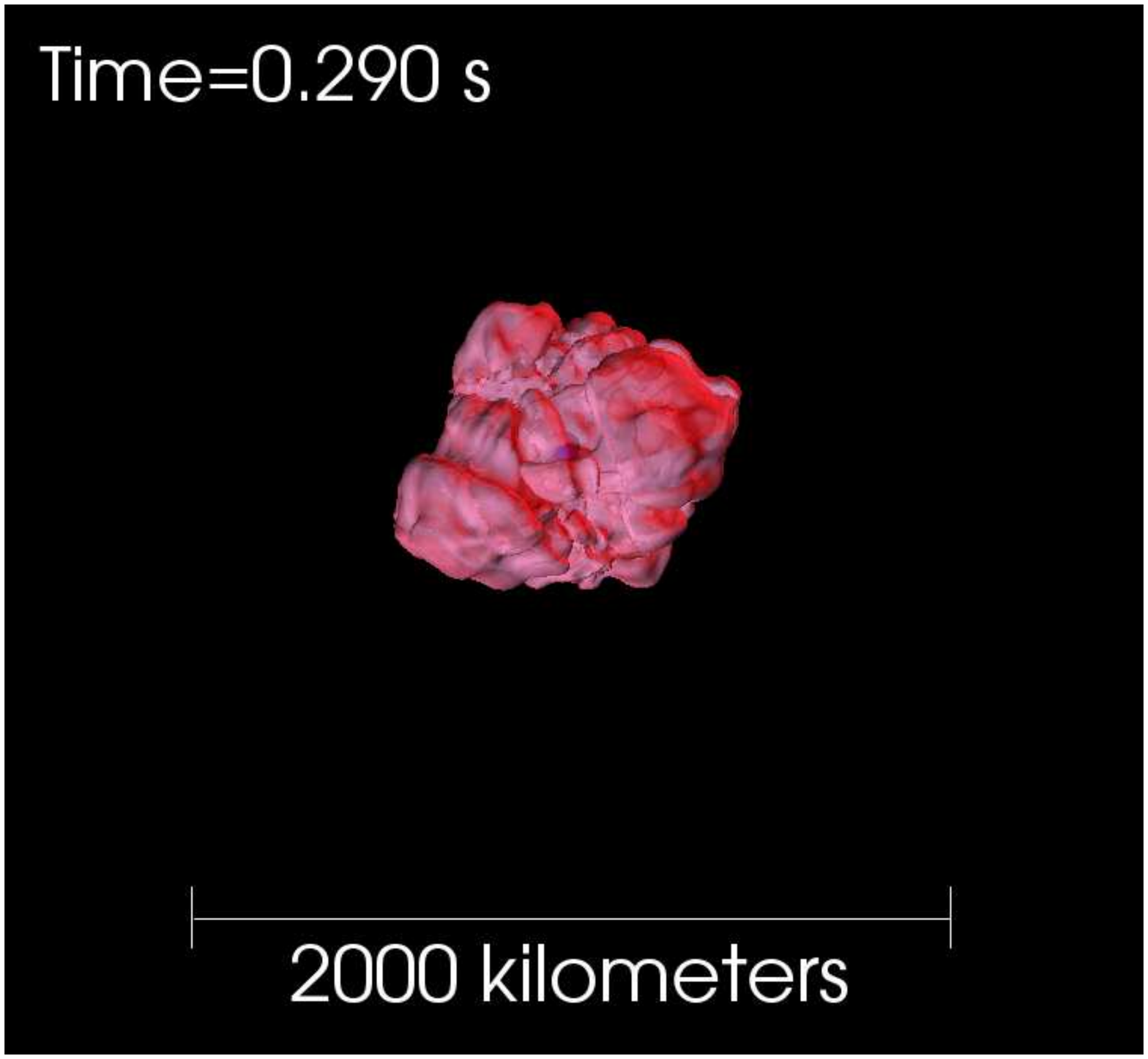}
\includegraphics[height=.5\textheight]{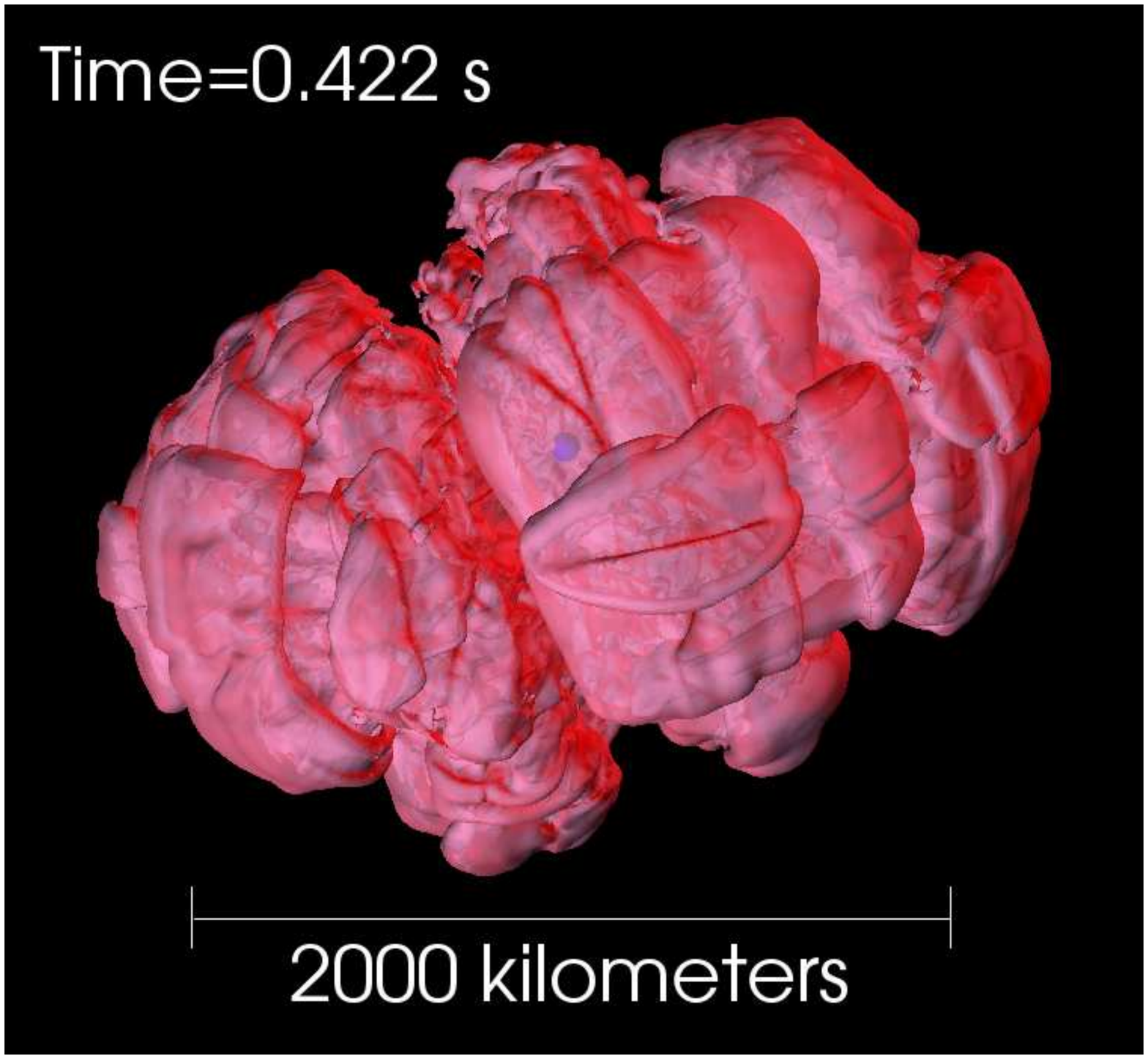}}
\caption{Blast Morphology. These panels depict an evolutionary sequence (top left, top right, bottom left,
bottom right) of the $L_{\nu_e} = 1.9\times 10^{52}$ erg s$^{-1}$
model in 3D.  The top left panel depicts a phase well before the onset of explosion.
The two surfaces in each panel are $\rho = 10^{12}$ g cm$^{-3}$ (blue interior) and Y$_e$ = 0.47 (outer) for four
different times after bounce (0.156, 0.201, 0.289, and 0.422 seconds).  The scale is more than 2000 kilometers on a side.
Note that the crude axis of the explosion is not along any of the
three Cartestian directions and that there is no obvious $\ell = 1$ asymmetry.
See \S\ref{conclusions} for a discussion of the possible implications of this figure and of
Fig. \ref{fig3} for the relevance of the $\ell = 1, m = 1$ ``SASI" mode 
in the context of core-collapse supernovae.}
\label{fig6}
\end{figure}

In addition, as Figs. \ref{fig3} and \ref{fig6} indicate, even though we
do see pre-explosion and blast asymmetries in 3D, we do not
see a strong $\ell = 1$ axially-symmetric ``SASI" oscillation,
such as is obtained in 2D in this paper and by others (Blondin et al.
2003; Bruenn et al. 2007,2010; Burrows et al. 2006,2007a; Buras et al.
2006ab; Marek \& Janka 2009).  Rather, the free energy available to power
instabilites seems to be shared by more and more degrees of freedom
as the dimension increases (Iwakami et al. 2008).  In 1D, as demonstrated by Murphy \& Burrows (2008),
one sees a strong radial oscillation when the luminosity is near critical.
However, 2D models do not manifest this radial mode, but instead execute
an $\ell = 1, m = 0$ dipolar oscillation.  In 3D, this dipolar oscillation exists
(Blondin \& Mezzacappa 2007; F{e}rnandez 2010), but in the linear limit competes with
the $m = \{-1, 1\}$ modes.  The $\ell = 1, m = 0$ mode is even less in
evidence in the non-linear limit, which is achieved early in our simulations. Hence, the strong
dipolar symmetry seen in 2D and identified as the fundamental characteristic of the ``SASI" may not
survive in 3D.  At the very least, the $\ell = 1, m = 0$ mode does not dominate in our non-rotating 3D
models.  Whether rotation can change this conclusion remains to be seen (Guilet, Sato, \& Foglizzo 2010).

\section{Conclusions}
\label{conclusions}

In this paper, we have performed 1D, 2D, and 3D numerical simulations
of the collapse, bounce, and explosion of a massive star core to isolate 
the effect of spatial dimension in the context of the neutrino heating 
mechanism of core-collapse supernova explosions.  We have found that both the
viability of explosion (measured by the driving neutrino luminosity needed
to overcome a given mass accretion rate at the shock and its associated accretion 
pressure tamp) and the position of the ``critical curve" 
are monotonically increasing functions of dimension, with 3D more viable 
than 2D by 15$-$25\% and 3D more viable than 1D by almost a factor of two.  
Some had thought that 3D would prove similar to 1D, but our calculations do not support 
this expectation.  Moreover, we have discovered that the time to explosion is 
significantly shorter in 3D than 2D, all else being equal. These results suggest
that a key missing ingredient in the recipe 
for this astrophysically central phenomenon has been access to  
3D simulation tools and that the neutrino-driven explosion
mechanism is fundamentally 3D.  Two-dimensional simulations have to date
yielded marginal explosions and/or ambiguous answers (for a discussion see Janka et al. 2007). 

Furthermore, we find that the prominent $\ell = 1, m = 0$ dipolar mode 
of shock oscillation, often identified in 2D as a central aspect of the ``SASI,"
is little in evidence in 3D.  The free energy available to excite hydrodynamic 
instabilities seems instead to be shared by more degrees of freedom, diminishing
the amplitude of this axisymmetric dipolar component.  Whether this conclusion
is a function of progenitor and/or rotation has yet to be determined, but
the preliminary indications are that this sloshing mode, so visible in 2D
simulations, does not survive as a central feature of core collapse when 
three spatial dimensions are allowed for.

We emphasize that in this study, in order to isolate the crucial effect of dimension,
we employed a very simple neutrino ``transport" approach.
This has enabled us to discover what we 
suggest is a key aspect of the core-collapse supernova phenomenon. However, detailed 
neutrino transport must be incorporated to determine the true systematics 
with stellar progenitor of the explosion energies, nucleosynthesis, residual
neutron star (and black hole) masses, pulsar kicks, and blast 
morphologies.  If we had included such transport at this 
stage in the development of computer hardware and the computational arts, 
a single 3D simulation would have required many, many years of continuous 
execution on the largest existing and available massively-parallel platforms.
As a result, the insights we have achieved via our more modest study might have been 
delayed or obscured for years. This highlights the continuing benefits of a 
multi-pronged, varied, and flexible strategy to address this most important, 
though refractory, astrophysical problem. 

\acknowledgments
The authors acknowledge fruitful past collaborations with, conversations with, 
or input from Jeremiah Murphy, Christian Ott, Louis Howell, Rodrigo Fernandez, 
Manou Rantsiou, Tim Brandt, Dave Spiegel, Eli Livne, Luc Dessart, Todd Thompson, Rolf Walder,
Stan Woosley, and Thomas Janka. They would also like to thank Hank Childs and the VACET/VisIt 
Visualization team(s) for help with graphics and with developing multi-dimensional
analysis tools.  J.N. and A.B. are supported by the Scientific Discovery through 
Advanced Computing (SciDAC) program of the DOE, under grant number DE-FG02-08ER41544,
the NSF under the subaward no. ND201387 to the Joint Institute for Nuclear Astrophysics (JINA, NSF PHY-0822648),
and the NSF PetaApps program, under award OCI-0905046 via a subaward 
no. 44592 from Louisiana State University to Princeton University.
Work at LBNL was supported in part by the SciDAC Program under contract DE-FC02-06ER41438.
The authors would like to thank the members of the Center for Computational 
Sciences and Engineering (CCSE) at LBNL for their invaluable support for CASTRO.
A.B. and J.N. employed computational resources provided by the TIGRESS
high performance computer center at Princeton University, which is jointly supported by the Princeton
Institute for Computational Science and Engineering (PICSciE) and the Princeton University Office of
Information Technology; by the National Energy Research Scientific Computing Center
(NERSC), which is supported by the Office of Science of the US Department of
Energy under contract DE-AC03-76SF00098; on the Kraken and Ranger supercomputers,
hosted at NICS and TACC and provided by the National Science Foundation through
the TeraGrid Advanced Support Program under grant number TG-AST100001; and on the Blue Gene/P at the
Argonne Leadership Computing Facility of the DOE.

\begin{table*}
  \begin{center}
   \caption{Explosion times and accretion rates.  The explosion time,  
$t_{\rm exp}$, is calculated when the average shock radius, $\langle R_{\rm  
shock}\rangle$, reaches 400 km and does not recede during subsequent  
evolution.  The accretion rate at the time of explosion is calculated  
just exterior to the shock where the infalling envelope is spherical,  
and is given by $\dot{M}_{\rm exp} = 4\pi\rho v r^2$.  A ``$-$" indicates  
the model did not explode during the allotted maximum 
simulation time (1.4 seconds).  The bolded times are those for models
which both exploded within 1.4 seconds and were calculated in 
either both 1D and 2D or both 2D and 3D. As this table indicates, there were models
that were calculated in 1D, 2D, and 3D, but the corresponding 1D models 
did not explode within 1.4 seconds.
\label{table2}}
  \begin{minipage}{140mm}
   \begin{tabular}{@{}ccccccccllllr@{}}
   \hline
   \\
  \vspace{0.02in}
$L_{\nu_e,\bar{\nu}_e}$ & Model & $t_{\rm exp}$  &  $\dot{M}_{\rm exp}$ & Model &$t_{\rm exp}$ &$\dot{M}_{\rm exp}$ &Model &$t_{\rm
exp}$ &$\dot{M}_{\rm exp}$\\
($10^{52}$ erg s$^{-1}$)& 1D& (ms)&($M_\odot$ s$^{-1}$) & 2D & (ms)& ($M_\odot$ s$^{-1}$)& 3D &(ms) &($M_\odot$ s$^{-1}$)\\
\\
\hline
\\
   1.7      &1d:L\_1.7&{\bf--} &-- & 2d:L\_1.7  & {\bf --} & --&  3d:L\_1.7   &{\bf 415} &0.28 &  \\
   1.8   && &  & 2d:L\_1.8& -- &--&  \\
   1.9 &1d:L\_1.9&{\bf --} & --& 2d:L\_1.9& {\bf 1045} & 0.14&  3d:L\_1.9    & {\bf 254} & 0.39 \\
   1.95 & & && 2d:L\_1.95& 444 & 0.24 & &\\
    2.0     &1d:L\_2.0&{\bf --} &-- &2d:L\_2.0 & {\bf 240} &0.29    && & & \\
   2.1    &1d:L\_2.1&{\bf--} &-- &2d:L\_2.1 & {\bf 229} & 0.33 & 3d:L\_2.1&{\bf 154}  & 0.56 &\\
   2.3    && & &  2d:L\_2.3       & 217 & 0.39 &\\
   2.5    &1d:L\_2.5& {\bf 748} &0.15 &  2d:L\_2.5      & {\bf 174} & 0.43 &\\
   2.6   &  1d:L\_2.6   & {\bf 679} & 0.18 & 2d:L\_2.6 &{\bf 165} & 0.48 \\
   2.7    &1d:L\_2.7& {\bf 568}& 0.21 &  2d:L\_2.7     & {\bf 163} & 0.49 &\\
   2.8    &1d:L\_2.8 & {\bf 365} & 0.27 &  2d:L\_2.8      & {\bf 155} & 0.53 &\\
   2.9    &1d:L\_2.9& 178  &0.41 &     &  &  &\\
   3.0   &  1d:L\_3.0   & 168 & 0.47 & &\\
   3.1   &  1d:L\_3.1   & 159 & 0.51 & &\\
  \\
  \hline
  \\
  \end{tabular}
\end{minipage}
\end{center}
\end{table*}

\newpage

\clearpage

\clearpage

\clearpage

\clearpage

\clearpage

\end{document}